\documentclass[aps,prl,showpacs,twocolumn,floats]{revtex4}
\usepackage{amssymb}

\usepackage[xdvi]{graphicx}

\begin{document}

\title{\textsc{Dynamic approach for micromagnetics close to the Curie
temperature}} 

\author{O.\ Chubykalo-Fesenko$^1$, U.\ Nowak$^2$, R.\ W.\
Chantrell$^2$, and D.\ Garanin$^3$} 

\affiliation{$^1$ Instituto de Ciencia de Materiales de Madrid, CSIC,
Cantoblanco, 28049 Madrid, Spain}

\affiliation{$^2$ Department of Physics, University of York, York
YO10 5DD, UK}

\affiliation{$^3$ \mbox{Department of Physics and Astronomy, Lehman
College, City University of New York,} \\ \mbox{250 Bedford Park
Boulevard West, Bronx, New York 10468-1589, U.S.A.}}
\date{\today}

\begin{abstract}
In conventional micromagnetism magnetic domain configurations are
calculated based on a continuum theory for the magnetization which is
assumed to be of constant length in time and space. Dynamics is
usually described with the Landau-Lifshitz-Gilbert (LLG) equation the
stochastic variant of which includes finite temperatures. Using
simulation techniques with atomistic resolution we show that this
conventional micromagnetic approach fails for higher temperatures
since we find two effects which cannot be described in terms of the
LLG equation: i) an enhanced damping when approaching the Curie
temperature and, ii) a magnetization magnitude that is not constant in
time.  We show, however, that both of these effects are naturally
described by the Landau-Lifshitz-Bloch equation which links the LLG
equation with the theory of critical phenomena and turns out to be a
more realistic equation for magnetization dynamics at elevated
temperatures.
\end{abstract}
 \maketitle




An increasing amount of research is focusing on the dynamic behavior
of ferromagnetic materials at elevated temperatures. The motivations
for this are manifold. A major imperative is the understanding of
pulsed laser experiments on thin film samples, for example the all
optical FMR experiments of Van Kampen et. al. \cite{vankampen}, and
the higher laser power experiments of Beaurepaire et. al.,
\cite{beaurepairePRL96} who demonstrated complete demagnetization on a
timescale of picoseconds. One of the main issues of the
high-temperature magnetization dynamics is the rate of the
magnetization relaxation due to different processes involving magnon,
phonon and electron interactions that contribute to thermal spin
disordering.

The basis of most of theoretical investigations of thermal
magnetization dynamics is a micromagnetic approach which considers the
magnetization of a small particle or a discrete magnetic nanoelement
as a vector of a fixed length (referred to here as a macro-spin) with the
phenomenological Landau-Lifshitz-Gilbert (LLG) equation of motion
augmented by a noise term \cite{lyberatosJPC93}. However, contrary to
atomic spins, there is no reason to assume a fixed magnetization
length for nanoelements at non-zero temperature. For instance, the
latter can decrease in time upon heating by a laser pulse. Hence, from
the point of view of modeling of magnetization dynamics, there is a
general need for further development of the micromagnetic theory in
terms of its ability to deal with elevated temperatures.

Within this context we note the failure of micromagnetics in general
to deal with the high frequency spin-waves which give rise to the
variation of magnetization with temperature. It has been suggested to
treat this problem using scaling approaches \cite{koch,vienna}. A
similar problem arises in multi-scale modeling (with atomistic and
micromagnetic discretizations to treat, for example, interfaces
\cite{felipe}) which can not correctly describe the transfer of high
energy spin-waves from atomistic into the micromagnetic region. An
alternative approach is the coarse graining model of Dobrovitksi
et. al. \cite{slav}, which has the advantage of being able to link the
length-scales but has been developed for simple systems only.

Some understanding of the pulsed laser experiments could indeed be
obtained in terms of a micromagnetic approach taking into account, in
an empirical way, the temperature variation of the intrinsic
parameters, particularly the saturation magnetization $M_s$ and the
anisotropy energy density $K$.  Lyberatos and Guslienko \cite{lg} have
used this macro-spin model to investigate the response of
nanoparticles during the Heat Assisted Magnetic Recording (HAMR)
process. The validity of the macro-spin approach including the thermal
variation of model parameters has further been investigated in Ref.\
\cite{uli:prb} using an atomistic approach. This work demonstrates
that, although the macro-spin model works well for temperatures far
below the Curie temperature $T_c$, longitudinal fluctuations of the
magnetization become important at elevated temperatures, which cannot
be treated within the macro-spin model of the corresponding LLG
equation of motion. The use of a macro-spin of fixed length places the
same physical constraint on micromagnetics at temperatures close to
$T_c$. Clearly, some approach to macro-spin dynamics beyond the LLG
equation is needed.

A semi-phenomenological equation of motion for macro-spins allowing for
longitudinal relaxation has been derived in Ref. \cite{Garanin} within
the mean-field approximation (MFA) from the classical Fokker-Plank
equation for individual spins interacting with the environment. This
``Landau-Lifshitz-Bloch (LLB) equation'' has been shown to be able to
describe linear domain walls, a domain wall type with non-constant
magnetization length. The validity of these results has been confirmed
by measurements of the domain wall mobility in YIG crystals close to
$T_{c}$ \cite{YIG} and by recent atomistic simulations
\cite{kazantsevaPRL05}.

In this letter we explore high-temperature dynamic properties using
atomistic modeling. These simulations are still based on the LLG
equation on the atomic level and, hence, do still not provide a
microscopic description of the damping itself. Nevertheless they do
include thermal degrees of freedom microscopically and encapsulate
important phenomena associated with relaxation, including the
thermodynamics of the phase transition and both, longitudinal and
transverse macroscopic relaxation. We find an enhanced transverse
relaxation when approaching the Curie temperature from below and a
magnetization magnitude which is not constant in time. Both of these
phenomena cannot be understood in terms of conventional micromagnetism
but, comparing these predictions with a macro-spin model based on the
LLB equation, we conclude that here these phenomena are indeed well
described by the LLB equation.

For our atomistic simulations we use a model in which the dynamic
behavior of classical spins $ |\mathbf{s}_{i}|=1$ on lattice sites $i$
with magnetic moment $\mu _{0}$ is treated at the atomic level with
the Langevin form of the LLG equation
\begin{equation}
  \dot{\mathbf{s}}_{i}=-\gamma \lbrack \mathbf{s}_{i}\times \mathbf{H}%
  _{i}]-\gamma \alpha \lbrack \mathbf{s}_{i}\times \lbrack \mathbf{s}%
    _{i}\times \mathbf{H}_{i}]]  \label{Langevin-LLG}
\end{equation}
where $\gamma $ is the gyromagnetic ratio, and $\alpha $ is the
damping parameter, $\alpha =0.1$ in our simulations. The total field
$\mathbf{H}_{i}$ contains nearest-neighbor Heisenberg exchange
(exchange constant $J$) and Zeeman contributions and it is augmented
by a white-noise field $\mathbf{\zeta }_{i}(t)$ with the correlator
$\langle \zeta _{i\mu }(t)\zeta _{j\nu }(t^{\prime })\rangle
={\frac{2\alpha k_{B}T}{\gamma \mu _{0}}}\delta _{ij}\delta _{\mu \nu
\ }\delta (t-t^{\prime }),$ where $\mu ,\nu =x,y,z$.  For simplicity,
the dipolar interaction is neglected as well as any crystalline
anisotropy. A cubic lattice with periodic boundary conditions and
system sizes of $48^{3}$ has been considered. In the calculations we
first establish thermal equilibrium for a given temperature starting
with all magnetic moments parallel to the $z$ axis and applying a
field $H_{z}=0.05J/\mu _{0}$. Then, to evaluate the transverse
relaxation, all spins were simultaneously rotated by an angle of
$30^{\circ }$. We have calculated the average spin polarization
$\mathbf{m}=(1/N)\sum_{i}\left \langle \mathbf{s}_{i}\right\rangle $
per lattice site which is proportional to the experimentally observed
magnetization $\mathbf{M}$.

\begin{figure}[tbp]
  \includegraphics[width=7.9cm]{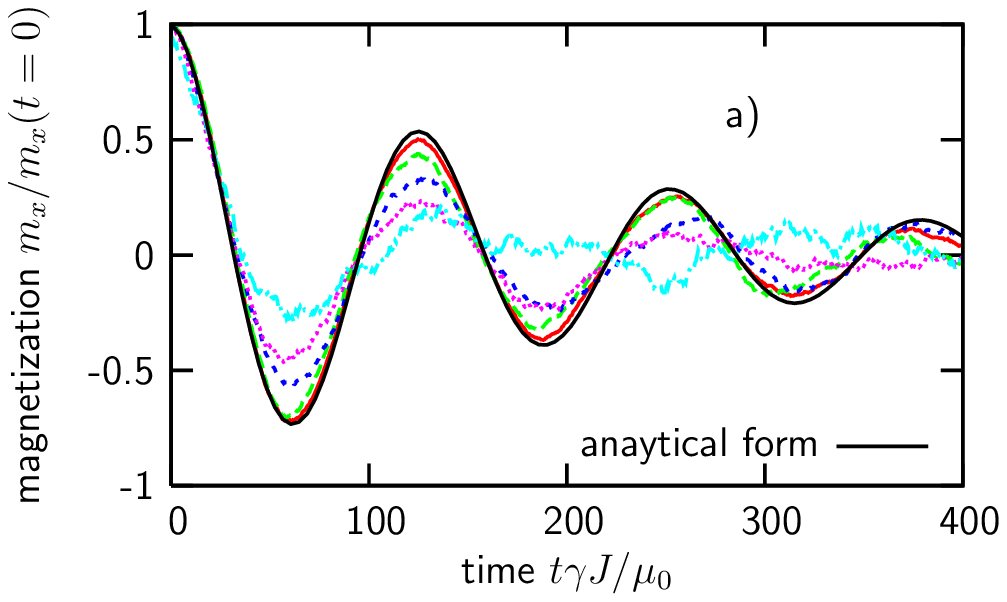}\newline
  \includegraphics[width=7.9cm]{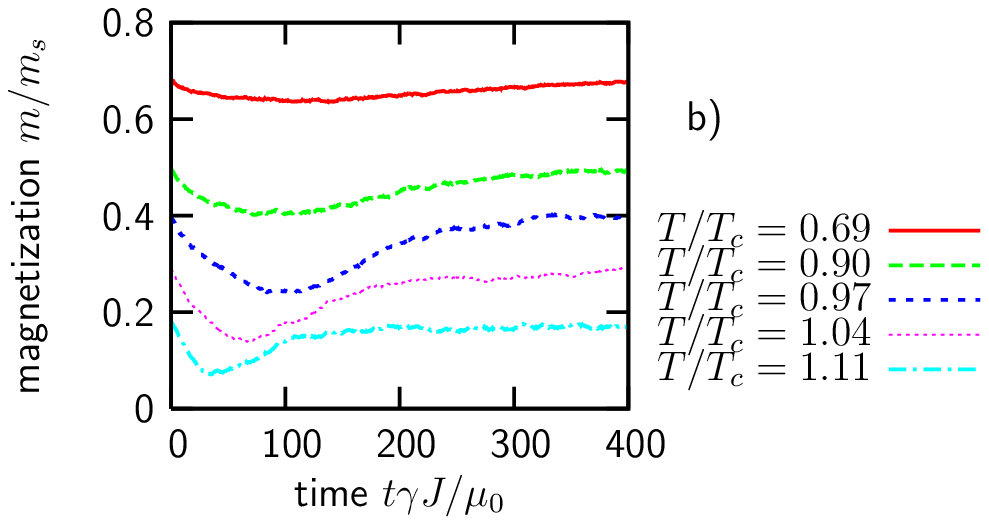}
  \caption{Relaxation of the magnetization for different temperatures
    using the atomistic modeling: a) normalized perpendicular
    component ($30^{\circ }$ excitation); b) absolute value of the
    magnetization $m\equiv \left| \mathbf{m}\right| $ ($135^{\circ
    }$ excitation).}
  \label{dyn_fit}
\end{figure}

Fig.~\ref{dyn_fit}a shows one transverse magnetization component as a
function of time for different temperatures. The magnetization is
normalized to its initial value and the data show clearly a faster
relaxation for higher temperatures. Note that in our simulation even
above the Curie temperature $T_{c}$ there is still a finite
magnetization due to finite-size effects and the fact that the
simulations are conducted within an external field. Fitting the curves
to an expression $m_{x}(t)\sim \cos (t/\tau _{p})\exp (t/\tau _{\perp
})$ shows a perpendicular relaxation time $\tau _{\perp }$ which
increases with temperature, deviating from its zero temperature limit
$1/(\alpha \gamma H_{z})$. Fig.\ \ref{dyn_fit}b presents the change of
the absolute magnetization value as a function of time for a similar
simulation but with a large angle of $135^{\circ }$ . Note that the
magnetization magnitude shows a dip during the relaxation process
which is well below its equilibrium value. A dynamic response of this
type cannot be described in terms of the macro-spin LLG equation which
conserves the absolute value of the magnetization, but is consistent
with the LLB equation as will be discussed below.

It is interesting to note that our atomistic model with a constant
microscopic damping parameter exhibits an increase in the effective
macroscopic damping as observed experimentally \cite{liJAP91}. We
believe that this is due to magnon-magnon scattering processes which
give rise to the initial decrease of the magnetization as the energy
is transferred from the $k=0$ (precessional) mode to higher order
modes. This results in the enhanced transverse damping and in the dip of
the magnetization, followed by the recovery to its equilibrium value
as the spin waves decay.

\begin{figure}[tbp]
  \includegraphics[angle=-90,width=9.cm]{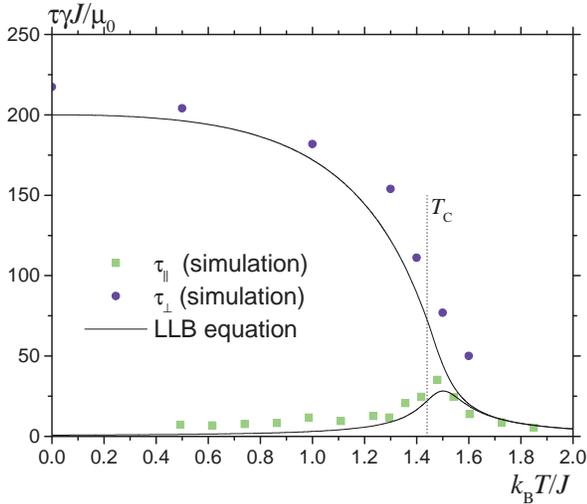}
  \caption{Temperature dependence of longitudinal and transverse
    relaxation times from the atomistic modeling and the LLB equation,
    calculated as inverse rates given by Eq.~(\ref{Gammas}). }
  \label{Slowing}
\end{figure}

Furthermore, we investigate the longitudinal relaxation time $\tau
_{\Vert }$ from the initial relaxation of the fully ordered system to
thermal equilibrium. The relaxation of the magnetization to
equilibrium is found to be approximately exponential on longer time
scales which defines the characteristic time $\tau _{\parallel
}$. Fig.\ \ref{Slowing} shows the variation of the longitudinal
relaxation time with temperature. The rapid increase close to $T_{c}$
is known as critical slowing down \cite{Landau}, a general effect
characterizing second order phase transitions. Also shown in Fig.\
\ref{Slowing} is the perpendicular relaxation time $\tau _{\perp }$
determined as described above. Approaching the Curie temperature the
perpendicular relaxation time $\tau _{\perp }$ breaks down.

As we have demonstrated so far, the atomistic model shows important
physical aspects of the behavior of nanoscale magnetic systems,
including a temperature dependence of the effective damping,
longitudinal fluctuations and critical slowing down. Next, we
demonstrate that these effects can be described alternatively by
macro-spin magnetization dynamics in terms of the
Landau-Lifshitz-Bloch equation of motion \cite{Garanin}. This provides
not only a deeper understanding of the phenomena but it also suggests
that the LLB equation is more suitable than the LLG equation for
finite temperature micromagnetics.

The LLB equation following from Eq. (\ref{Langevin-LLG}) in the
spatially homogeneous case can be written in the form
\begin{eqnarray}
  &&\mathbf{\dot{m}}=-\gamma \lbrack \mathbf{m}\times \mathbf{H}_{\mathrm{eff}%
    }]+\gamma \alpha _{\parallel }\frac{(\mathbf{m\cdot H}_{\mathrm{eff}})%
      \mathbf{m}}{m^{2}}  \nonumber \\
    &&\qquad {}-\gamma \alpha _{\perp }\frac{[\mathbf{m}\times \lbrack \mathbf{m}%
	\times \mathbf{H}_{\mathrm{eff}}]]}{m^{2}},  \label{LLBm}
\end{eqnarray}
where $\mathbf{m=\langle s\rangle }$ is the spin polarization and
$\alpha _{\parallel }$ and $\alpha _{\perp }$ are dimensionless
longitudinal and transverse damping parameters given by
\begin{equation}
  \alpha _{\parallel }=\alpha \frac{2T}{3T_{c}^{\mathrm{MFA}}},\quad \alpha
  _{\perp }=\alpha \left[ 1-\frac{T}{3T_{c}^{\mathrm{MFA}}}\right]
  \label{alphasClass}
\end{equation}
for $T<T_{c}^{\mathrm{MFA}}$ and the same with $\alpha _{\perp
}\Rightarrow \alpha _{\parallel }$ for $T>T_{c}^{\mathrm{MFA}},$ where
$T_{c}^{\mathrm{MFA}}$ is the mean-field Curie temperature.  Here,
$\alpha $ is the same damping parameter that enters
Eq. (\ref{Langevin-LLG})$.$ The effective field $\mathbf{H}_{%
\mathrm{eff}}$ is assumed to be much weaker than the exchange
interaction and it is given by
\begin{equation}
\mathbf{H}_{\mathrm{eff}}=\mathbf{H}+\mathbf{H}_{A}+\left\{
\begin{array}{cc}
\frac{1}{2\tilde{\chi}_{\Vert }}\left( 1-\frac{m^{2}}{m_{e}^{2}}\right)
\mathbf{m,} & T\lesssim T_{c}^{\mathrm{MFA}} \\
\frac{J_{0}}{\mu _{0}}\left( \epsilon -\frac{3}{5}m^{2}\right) \mathbf{m,} &
T\gtrsim T_{c}^{\mathrm{MFA}}
\end{array}
\right. \mathbf{.}  \label{Heffm}
\end{equation}
Here $\mathbf{H}$ and $\mathbf{H}_{A}$ are applied and anisotropy
fields and $m_{e}$ is the zero-field equilibrium spin polarization in
the MFA that satisfies the Curie-Weiss equation
\begin{equation}
  m=B\left[ \beta (mJ_{0}+\mu _{0}H)\right]   
  \label{Curie-Weiss}
\end{equation}
with $H=0$ and $\epsilon \equiv 1-T/T_{c}^{\mathrm{MFA}}$. $B$ is the
Langevin function, $\beta =1/\left( k_{B}T\right) ,$ and $J_{0}$ the
zero Fourier component of the exchange interaction related to
$T_{c}^{\mathrm{MFA}}$ as $k_{B}T_{c}^{\mathrm{MFA}}=J_{0}/3.$ In
Eq. (\ref{Heffm}) $\tilde{\chi}_{\Vert }=\partial m(H,T)/\partial H$
is the longitudinal susceptibility at zero field that can be obtained
from Eq. (\ref{Curie-Weiss}). The anisotropy field $\mathbf{H}_{A}$
due to the uniaxial anisotropy is related to the zero-field transverse
susceptibility $\tilde{\chi}_{\perp }$ as $\mathbf{H} _{A}=\left(
m_{x}\mathbf{e}_{x}+m_{y}\mathbf{e}_{y}\right) /\tilde{\chi} _{\perp
}$\cite{Garanin}$.$ The equilibrium solution of the LLB equation
satisfies $\mathbf{m}\times \mathbf{H}_{\mathrm{eff}}=0$ and
$\mathbf{m\cdot H}_{\mathrm{eff}}=0.$ For $T\ll T_{c}^{\mathrm{MFA}}$
the longitudinal susceptibility $\tilde{\chi}_{\Vert }$ becomes very
small in which case it can be shown that $m\cong m_{e}.$ This means
that the longitudinal relaxation vanishes and Eq. (\ref{LLBm}) reduces
to the standard LLG equation with $ \alpha _{\bot }=\alpha .$

In the damping parameters $\alpha _{\parallel }$ and $\alpha _{\perp
}$ of Eq. (\ref{alphasClass}) $\alpha $ is non-critical at
$T_{c}^{\mathrm{MFA}}.$ Its temperature dependence cannot be
established within our semi-phenomenological approach, so we assume it
to be a constant, for the sake of comparison with the results of our
atomistic simulations. 
The LLB equation also can be written in terms of the vector
$\mathbf{n}= \mathbf{m}/m_{e}$ \cite{garchub04prb}. This form
provides a link to the micromagnetic anisotropy constants but becomes
inconvenient above $T_{c}$ where $m_{e}$ disappears.

In order to effect a comparison we analyse the relaxation rates
derived from the LLB equation. Firstly we note from Eq.\
\ref{alphasClass} a linear increase of $\alpha_{\parallel }$ with $T$,
whereas the behavior of $\alpha _{\perp }$ is non-monotonic, changing
from a linear decrease below $T_{c}^{\mathrm{MFA}}$ to a linear
increase above $T_{c}^{\mathrm{MFA}}$. However, it is important to
note that $\alpha_{\parallel }$ and $\alpha _{\perp }$ are
non-critical for all finite temperatures, and that the variation of
$\alpha _{\perp }$ is weak. With this background, we now consider the
relaxation rates from the linearized LLB equation which have the form
\begin{equation}
  \Gamma _{\parallel }=\frac{\gamma \alpha _{\parallel
    }}{\tilde{\chi}_{\Vert }(H,T)},\qquad \Gamma _{\perp
    }=\frac{\gamma \alpha _{\perp }}{\tilde{\chi}_{\perp }(H,T)},
  \label{Gammas}
\end{equation}
where $\tilde{\chi}_{\Vert }(H,T)$ is the longitudinal susceptibility
at nonzero field that follows from Eq. (\ref{Curie-Weiss}) or simply
from $\mathbf{m\cdot H}_{\mathrm{eff}}=0,$ in our approximation. 

The longitudinal relaxation rate is, in general, very fast as $\Gamma
_{\parallel }\sim J_{0}$. Since $\tilde{\chi}_{\Vert }(H,T)$ is large
near $T_{c}^{\mathrm{MFA}},$ $\Gamma _{\parallel }$ shows critical
slowing down which is a result of the critical behavior of
$\tilde{\chi}_{\Vert }(H,T)$ rather than the variation of $\alpha
_{\parallel}$.  The transverse susceptibility for the isotropic model
is simply given by $\tilde{\chi}_{\perp }(H,T)=m(H,T)/H$ so that
$\Gamma _{\perp }\sim H$ is much smaller than $\Gamma _{\parallel }$
below $T_{c}^{\mathrm{MFA}}$. However, it increases with temperature,
as was observed in the atomistic modeling presented above and its
critically behavior close to $T_c$ is $\Gamma _{\perp }\sim
1/m(H,T)$. For temperatures below $T_c$ a corresponding behavior was
found for the line widths of FMR experiments \cite{liJAP91}.

At $T=T_{c}^{\mathrm{MFA}}$ the rates are given by
\begin{equation}
  \Gamma _{\parallel }\cong \frac{6}{5}\frac{\gamma \alpha J_{0}}{\mu
    _{0}} m_{H}^{2},\qquad \Gamma _{\perp }\cong
  \frac{2}{5}\frac{\gamma \alpha J_{0}}{\mu _{0}}m_{H}^{2}
  \label{Gamma1Tc}
\end{equation}
where $m_{H}=\left[ \left( 5/3\right) \left( \mu _{0}H/J_{0}\right) \right]
^{1/3}$ is the induced magnetization at $T_{c}^{\mathrm{MFA}}.$ Above $%
T_{c}^{\mathrm{MFA}}$ both rates merge:
\begin{equation}
  \Gamma _{\parallel }\cong \Gamma _{\perp }\cong
  \frac{2}{3}\frac{\gamma \alpha J_{0}}{\mu
    _{0}}\frac{T}{T_{c}^{\mathrm{MFA}}}\left( \frac{T}{T_{c}^{
      \mathrm{MFA}}}-1\right) .  
  \label{GammasAbove}
\end{equation}
Finally, in the presence of uniaxial anisotropy $\Gamma _{\perp }$ is
given by Eq. (\ref{Gammas}) with $1/\tilde{\chi}_{\perp
}(H,T)=H/m(H,T)+ 1/\tilde{\chi}_{\perp }$, where $\tilde{\chi}_{\perp
}$ is only weakly temperature dependent within mean-field theory below
$T_{c}^{\mathrm{MFA}}$.

\begin{figure}[tbp]
  \includegraphics[width=7.9cm]{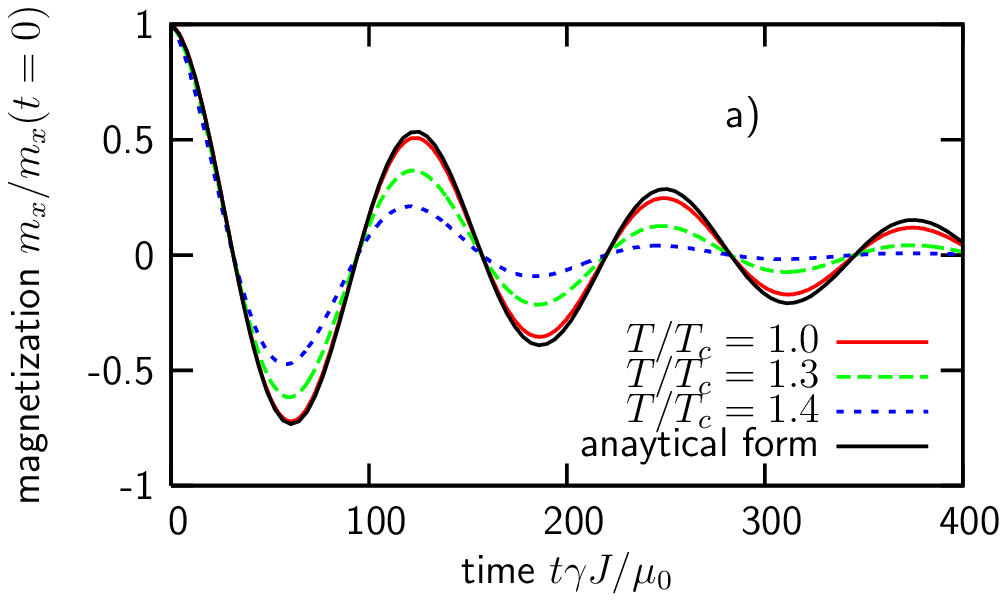}\newline
  \includegraphics[width=7.9cm]{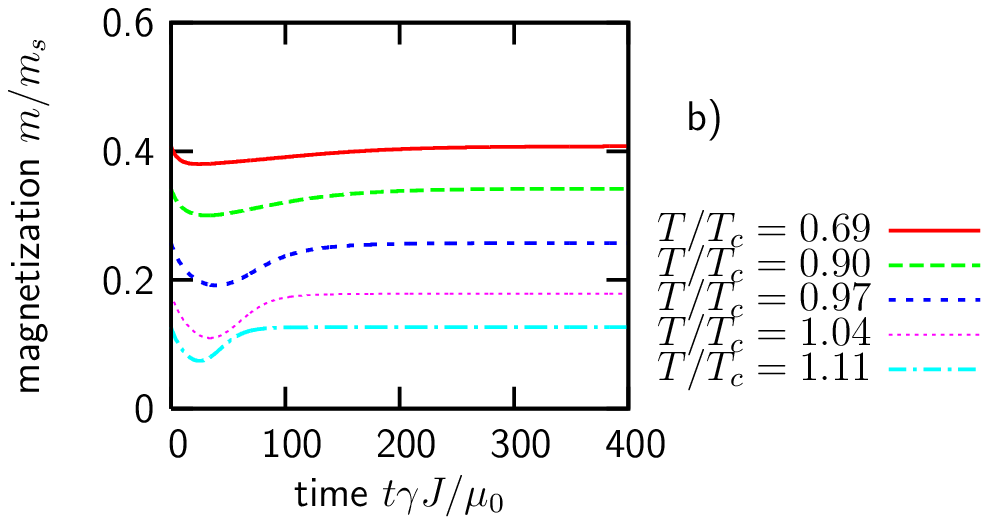}
  \caption{Relaxation of the magnetization for different temperatures as in
    Fig.\ \ref{dyn_fit} but using the macro-spin LLB modeling.}
  \label{MF_mod}
\end{figure}

To compare the LLB results with the predictions of the atomistic
model, Fig. \ref {Slowing} includes the inverse relaxation rates
calculated using Eq. (\ref{Gammas}) with rescaled temperature to fit
the exact value $k_{B}T_{c}=1.44J$ for a simple cubic lattice. The
agreement between Eq. (\ref{Gammas}) and the numerical results is
remarkable given the MFA used in the derivation of Eq. (\ref{Gammas}).

Also, we have integrated numerically Eq.\ (\ref{LLBm}) for a
macro-spin to give the time evolution of the magnetisation components
for comparison with the numerical results of Fig.\ \ref{dyn_fit}. The
results are presented in Fig.\ \ref{MF_mod}. Comparison with Fig.\
\ref{dyn_fit} shows that the LLB equation reproduces essential
physical processes which govern the magnetization dynamics at elevated
temperatures and thus it can be used as an alternative to
micromagnetics in this region.  However, this comparison could still
be improved if one evaluates the macro-spin parameters directly from
an atomistic simulation.  Furthermore, if the LLB equation is to be
used as an alternative to micromagnetics, the corresponding parameters
could as well be extracted from experiment.

In conclusion, performing atomistic simulations of thermal
magnetization dynamics we observe an increase of the macroscopic
transverse \ damping approaching the Curie temperature. This increase
is determined by the thermal dispersion of magnetization and would
exist independently from any other possible thermal dependence of
internal damping mechanisms such as phonon-magnon coupling. This
effect explains the broadening of the resonance line width in
classical FMR experiments \cite{liJAP91}. Furthermore, the
magnetization vector turns out not to be constant in length. Instead
during relaxation one can observe a dip of the magnetization which is
more pronounced when approaching the Curie temperature. Finally, the
magnetization dynamics has important contributions from longitudinal
relaxation. This relaxation shows critical slowing down at
temperatures close to $T_{c}$. Importantly, the observed dynamics is
in agreement with the dynamics of a macro-spin described by the
Landau-Lifshitz-Bloch equation which contains both longitudinal and
transverse relaxation. This equation could serve in future as a basis
for an improved micromagnetics at elevated temperature.

\acknowledgments{This work was supported by a joint travel grant of
the Royal Society (UK) and CSIC (Spain).}


\end{document}